%% file: ms_preprint.tex
\documentclass[preprint]{aastex}

\newcommand{\hi}{{H~{$\scriptstyle {\rm I}$}}}

\begin{document}
\title{Discovery of a Population of \hi\ Clouds in the Galactic Halo \\  
{\it (Accepted for Publication in the Astrophysical Journal Letters) \\ }
{\it (Preprint -- October 14, 2002)} }

%% revised version 08 October 2002
 
\author{Felix J. Lockman}
\affil{National Radio Astronomy Observatory
\footnote{The National Radio Astronomy Observatory is a facility of the
National Science Foundation operated under cooperative agreement with
Associated Universities, Inc.}, P.O. Box 2, Green Bank, WV, 24944; 
 jlockman@nrao.edu}
\authoraddr{P.O. Box 2, Green Bank, WV, 24944}

\begin{abstract}

A population of discrete \hi\ clouds in the halo of the inner 
Galaxy has been discovered in 21cm observations made with the 
Green Bank Telescope.  
 The halo clouds are seen up to  1.5 kpc from the 
Galactic plane at tangent points throughout the first longitude quadrant  
and at several locations in the fourth quadrant.  
Their velocities follow Galactic rotation.  
A group of clouds  more than 500 pc 
below the plane near  $\ell = 29\arcdeg$ was studied in detail.  
In the median, the 38 clouds  have a peak $N_{HI}$ of 
a few  $10^{19}$ cm$^{-2}$, a diameter of a few 
 tens of pc,  an \hi\ density of a few tenths cm$^{-3}$, and
an \hi\ mass of 50 $M_{\Sun}$, with a considerable range 
about the median.   Some halo clouds have line 
widths so narrow that their temperature must be $< 1000 $ K. Some  
appear to have a core-halo velocity structure.  As much as half the 
mass of the neutral halo may be in clouds.

\end{abstract}
 

\section{Introduction}

     The Galactic \hi\ halo --- the neutral gas layer which 
extends  many scale
heights above  the disk --- has been detected in 21cm emission, 
Ly$\alpha$ absorption, and 
in other species  (see the review by  \citet{savage95}).  
In the inner Galaxy, where 
tangent-point observations can locate the gas accurately, 21cm \hi\ emission 
has been found to  $|z| \geq 500 - 1000$ pc from the Galactic plane 
(\citet{l84}; hereafter L84), and combined 21cm-Ly$\alpha$ studies 
give  similar results for gas near the Sun \citep{lhs}. 
The spatial structure of halo \hi\ has always been uncertain.  
A list of ``halo clouds" was given in L84, but these clouds are quite large, 
possibly atypical,  and not obviously 
related to the more pervasive halo medium which appears 
 smooth, though lumpy, in the major 21cm surveys 
\citep{ww73, hartmann}.   A ``patchiness factor" for halo gas was introduced by
Savage and collaborators \citep{savage90, diplas94} 
 to account for the relatively large scatter 
in the absorption column densities from one direction to another, 
but targets for absorption studies are sparsely distributed across the
sky, and  give little information on the angular structure 
of  the  gas.  When 21cm survey spectra 
are averaged over large areas to achieve the highest 
sensitivity, they show what appears to be 
very high velocity-dispersion wings,  suggesting that the Galaxy has 
 an \hi\ component with a scale-height of 4 kpc \citep{kalberla98, 
kalberla02}. However, the large areal averaging necessary to 
detect this emission  removes most information on its angular structure.

     This Letter reports results from the first observations 
with the Robert C. Byrd Green 
Bank Telescope (the GBT) of 21cm emission from the Galactic \hi\ 
halo at the tangent points in
the inner Galaxy.  The GBT has good angular resolution, 
excellent sensitivity, and high
dynamic range for this work, and has revealed 
unexpected structure in  the \hi\ far from the Galactic plane.

\section{Observations}

Observation were made in several sessions from   April to 
 July 2002 using the GBT at the NRAO in Green Bank, WV.   
The telescope's 100 meter
diameter, unblocked  aperture  gives it an exceptionally clean 
main beam with a half-power
 width at 21 cm of $9\farcm2$, and a  first sidelobe 30 dB 
below the main beam \citep{fjlspie98, jewellspie, maddalenaspie}.  The
receiver system temperature on cold sky was about 20 K.  The 
detector was the NRAO
spectral processor, a Fourier Transform instrument with an 
efficiency near unity. The spectra have  a velocity
coverage of $\pm250$ km s$^{-1}$ around $+50$ km s$^{-1}$  LSR in 
512 channels, each with a 
velocity width of 1.25 km s$^{-1}$ and a spacing of 1.03 km  s$^{-1}$. 
Residual instrumental baselines are 
modeled adequately with a first or second order polynomial.  
%[xxx Observations were made in several sessions at a variety of hour
%angles to confirm the reality of these lines.]

\section{HI Clouds far from the Galactic Plane}

At the tangent points in the inner Galaxy, Galactic rotation is 
projected entirely along the line of sight, 
so gas at the ``terminal'' velocity, $V_t$, has a distance 
from the Sun, {\it r}, a distance from the Galactic center, ${\it R}$,
and a distance from the Galactic plane, {\it z}, 
which is known with uncommon accuracy, 
 at least for $R>2$ kpc  (L84, \citet{liszt84, burton92}; in this 
Letter the Sun-center distance $R_0 \equiv 8.5$ kpc). 
In the halo, gas at the tangent point which corotates with gas in the 
disk will appear at $V_t \ cos(b) \approx V_t$ for the latitudes 
studied here.  The intrinsic velocity dispersion of the ISM will 
produce some emission at velocities beyond $V_t$, but 
any material which lags significantly behind Galactic rotation 
will have  $|V_{LSR}| < |V_t|$.  A lagging halo will be displaced 
away from the terminal velocity and will not be seen in this analysis. 

Figure 1 shows \hi\ emission observed with  the GBT in a  slice through 
the Galactic plane at  longitude $16\arcdeg$.   
The terminal velocity is marked with a vertical
line at $+135$ km s$^{-1}$ (here $V_t$ is derived from the low-latitude 
$^{12}CO$  studies of  \citet{bg76}, \citet{clemens}, and \citet{dame01}).
 Most of the \hi\ at the highest velocities is concentrated 
to $|b|\leq 1\arcdeg$, but near latitude $+3\arcdeg$,
just beyond the terminal velocity,   there is a narrow line  component 
 at $V_{LSR}=145$ km s$^{-1}$   which lies quite a bit above the 
disk emission.  Subsequent maps made around this feature show that it is 
 a discrete \hi\ cloud 
which has an angular size $\sim 15\arcmin$ and  a peak $N_{HI} = 1.2 
\times 10^{19} $ cm$^{-2}$.  Its velocity places it at
 or near the tangent 
point, implying that it has $z=430$ pc,  a linear extent of 36 pc, 
and an \hi\ mass  $\approx 70 \ M_{\Sun}$.  The 21cm line is narrow, with 
a $\Delta v = 6.7 $ km s$^{-1}$ (FWHM).  Another small cloud 
 at $b = 2\fdg2$ and $V_{LSR} = +133 $ km s$^{-1}$ has similar properties 
and is +310 pc above the plane.

 \hi\ clouds like these have been detected  with the GBT 
at longitudes $341\arcdeg,$  $19\arcdeg,$ 
$25\arcdeg,$ $29\arcdeg,$ $35\arcdeg,$ $40\arcdeg,$ and $45\arcdeg$. 
Some of the brighter  halo clouds  
 can be recognized in low-resolution survey data 
(e.g., \citet{hartmann}), but they usually 
appear there in only one or two pixels.  These clouds 
constitute a previously undiscovered  population.  They are found  
 many degrees and many hundreds of pc from the Galactic plane.
They  appear to share the same general rotation as gas in the disk --- 
a Galactic rotation curve derived from the halo clouds alone 
would be quite similar to a rotation curve derived from the disk.  
Although the halo  clouds contain only a minuscule fraction 
of the total $N_{HI}$ in their direction 
(the total $N_{HI}$  is dominated by the disk), 
they may contain a significant fraction of the mass of the neutral halo.

\section{The \hi\ Halo  at Longitude $29\arcdeg$}

An area of the sky near $\ell = 29\arcdeg$
was mapped with the GBT  to provide information on a set of these 
objects at one location in the Galaxy ($R = 3.7$ kpc; $r = 7.6$ kpc). 
This region was known from  preliminary observations to have
 a number of tangent-point clouds, and it does not appear to 
have high optical extinction, so there is some hope that objects found 
in this field may be studied at other wavelengths.
Figure 2 shows \hi\  maps  averaged over 
 several velocity intervals.  The left panel covers emission 
 $5 - 15$ km s$^{-1}$ greater than $V_t$, which is about 
105 km s$^{-1}$ at this longitude \citep{clemens}.  
The map is dotted with clouds, so common near the plane that 
they blend into structures which are difficult to untangle.  
The central panel of Figure 2  shows the same area at 
 higher velocities.  There are  additional clouds, though 
fewer than at the lower velocities.  
The right panel, covering very high velocities, shows mainly noise and the 
wings of lower-velocity clouds, but also a
weak emission feature near $\ell,b = 29\fdg65 - 8\fdg9$, which arises in 
a small cloud at $V_{LSR} = 160$ km s$^{-1}$ 
that has a peak   $N_{HI} = 5 \times 10^{18}$ cm$^{-2}$ at $z = -1160$ pc.

\subsection{Observations of Individual Clouds}

More sensitive spectra of several minutes duration were obtained toward the 
centers of 38 bright, unconfused clouds chosen from preliminary 
observations of the $\ell = 29\arcdeg$ field. 
Gaussian components were fit to the spectra to determine the 
peak line intensity, $T_L$, the line width (FWHM) $\Delta v$,  and the 
velocity of each cloud.  The clouds all have $V_{LSR} \gtrsim V_t$ and 
$ b \leq -4\fdg3,$ and are thus at least  500 pc  from the plane.  
Their emission is typically five to ten times the background. 
  Properties 
of the 38 clouds are summarized in Table 1 and representative 
spectra are shown in Figure 3. 

In Table 1 
the quantity $N_{HI}$ is the column density at the cloud center, 
where $N_{HI} = 1.94 \times 10^{18}\  T_L\  \Delta v$ cm$^{-2}$. 
 The diameter listed is the square root of the 
product of the diameters in longitude and latitude.  
 About 25\% of the clouds are 
unresolved to the $9\arcmin$ beam of the GBT and most of the others 
are not greatly elongated.  
The average volume density, $\langle n \rangle$, is 
 $N_{HI}$ divided by the diameter, and thus  a lower limit for 
unresolved clouds.  Most of the clouds have $\langle n \rangle < 0.4$ 
cm$^{-3}$, but some have up to three times this value.  
The total mass is for \hi\ only.   The mass distribution is 
strongly skewed toward lower mass clouds:  one-third have 
$M_{HI} \leq 30 M_{\Sun}$.

\subsubsection{Velocity}

The 38 clouds have velocities in the range $103 - 148$ km s$^{-1}$, but the 
number per unit velocity decreases approximately linearly with 
$V_{LSR}$: half of the clouds lie within 15 km s$^{-1}$ of $V_t$.  
The total $N_{HI}$ in clouds also decreases with increasing velocity.
The clouds are thus related to the disk dynamically.  
  A simulation using a flat Galactic rotation curve suggests that a cloud 
population with a line-of-sight velocity dispersion 
$\sigma_v \approx 15 - 20$ 
km s$^{-1}$ would be consistent with the distribution of cloud velocities 
at $V_{LSR} \gtrsim V_t$.  There is no strong correlation of 
$V_{LSR}$ with $z$: the highest velocity clouds are found near 
$z \sim -1$ kpc.

\subsubsection{Kinetic Temperature and Line Shape}

Line widths are approximately uniformly distributed between 4.2 
and 20 km s$^{-1}$ with only a few  higher values.  
About one-quarter of the lines are so narrow that  
the clouds must be rather cool, with $T < 2000$ K, and   there 
are 4 clouds with $T < 1000$ K.  Some halo clouds listed in L84 
are also this cool.  It may be significant that the narrowest lines, 
those with $\Delta v < 9.0 $ km s$^{-1}$, 
 are all found in the half of the sample closest to the Galactic plane.
Virtually all of the lines from the 
clouds are superimposed on 
weak emission which often extends to higher  velocity $(\S 4.1.5)$. 
 Some  clouds appear to have a core-halo structure 
with two spectral components, one broad and one narrow, 
at the same position and velocity.  An example is 29.12-5.65 in Figure 3.
  The broad  component is typically one-third as bright as the 
narrow component, and has at least three times the line width, but 
the Gaussian decomposition is often 
ambiguous and properties of the spectral ``halos'' are quite uncertain.  
In particular, the observations are not yet sensitive enough to determine if
the apparent halos are conterminous with the cores. 

\subsubsection{Thermal Pressure}
An estimate of a cloud's thermal pressure is 
 $P_t \equiv nT \leq 22 \langle n \rangle \Delta v^2$. 
 In terms of observables, $P_t \propto T_L \Delta v^3 
Diam^{-1}$.  Values of $\Delta v$  are almost certainly dominated 
by turbulence and thus give only an upper limit on $P_t$. 
Also,  most of the clouds are no more than a few 
beamwidths across, so there is significant uncertainty in their diameter. 
  Estimates of $P_t$ from these data are thus
highly uncertain, but for what it is worth, 
clouds closer to the Galactic plane have $P_t$ typically half 
that of those farther from the plane, and there are 
 six clouds which appear to be resolved in angle and have  $P_t < 1000$ 
K-cm$^{-3}$.  Two-thirds of the clouds 
with the lowest $P_t$ seem to have core-halo spectral structure, where 
$P_t$ refers to the line from the core.  
Given the size and \hi\ mass of the clouds, their pressures, whether
thermal or turbulent, are orders of magnitude higher than 
the limit for gravitational binding, suggesting that the clouds, 
if they are stable, are confined by an external medium.  

\subsubsection{Correlations with Position and Velocity}

Linewidth is the only property of the halo clouds seemingly 
 correlated with distance from the Galactic 
plane: the narrower lines tend to lie closer to the  plane 
than the broader ones.  Cloud size, mass, and $N_{HI}$ are all 
independent of $z$.  

There is a relationship between $T_L$  and
$V_{LSR}$ in that the brightest lines come from clouds 
whose velocities are closest to $V_t$. 
The median value of $T_L$ for clouds at  $V_{LSR}$ $ < 110$ km s$^{-1}$
is more than twice that for clouds at  $V_{LSR} > 130$ km s$^{-1}$. 
%The twelve clouds with $V_{LSR}$ $ < 110$ km s$^{-1}$ 
%have a median $T_L$ of 1.7 K while the nine at $V_{LSR} > 130$ km s$^{-1}$ 
%have a median of 0.7 K.  
Line widths and cloud size have no 
strong relationship with $V_{LSR}$, but the decline in mean $T_L$ with 
increasing $V_{LSR}$ 
produces a nearly linear decline in the average $log(N_{HI})$ 
with $V_{LSR}$ from a value near 19.5 for clouds at $V_{LSR} = 110$
 km s$^{-1}$ to  19.0 at $V_{LSR} = 145$ km s$^{-1}$.  
Cloud diameters are not dependent on   $V_{LSR}$, 
but from the $N_{HI}(V_{LSR})$ trend,  it follows 
that the most massive clouds (and those with the largest $\langle n \rangle$)
 are found preferentially at the lowest velocities.

\subsubsection{The Fraction of the \hi\ Halo Contained in Discrete Clouds}

In addition to the clouds, there is a  broader component of 
\hi\ emission on which the clouds are superimposed both in velocity 
(Fig. 3) and position (Fig. 2). 
This emission may be from  faint, unresolved clouds, or a
genuinely diffuse medium.  
The 38 clouds contain about 20\% of the total $N_{HI}$ at $V_{LSR}>V_t$ 
within $-9\fdg0 \leq b \leq -4\fdg2$,  
though the amount varies from zero at some latitudes to $>80\%$ at others.
Considering that the 38 clouds studied in detail 
are only a subset of all the clouds 
visible in this field, it is plausible that 
 half of the \hi\ emission  in the $\ell=29\arcdeg$ 
field might come from  clouds. 
Because of velocity crowding, emission at a common $V_{LSR}$ may arise
from unrelated material separated by 
as much as a few kpc along the line of sight around the tangent point.
 From the current data it is not possible to determine if the 
halo clouds are related to the diffuse emission, or even if they 
are located in the same region of space.

\section{Summary Comments}

Observations with the GBT show that the halo of the inner Galaxy contains 
a population of \hi\ clouds  related dynamically to 
the disk.   This gas was already known to be in the halo from 
low-resolution studies (L84, \citet{dl90}); what was not known is that 
a significant fraction of it is  organized into clouds. 
 It is not yet possible to derive a scale height 
for the clouds, but by traditional standards some of them are extremely 
far from the Galactic plane.  It is tempting to speculate 
that they are a distant analog to the intermediate velocity clouds
seen in the vicinity of the Sun.  
Halo clouds are found in GBT data throughout the inner Galaxy except at 
a few directions at $\ell \lesssim 10\arcdeg$, but it is too early to 
draw conclusions about their overall distribution. 
These clouds are so ubiquitous that some are likely to 
 occult  extragalactic objects, allowing 
their abundances to be determined through optical or UV 
 spectroscopy.

Several theoretical studies have addressed the issue of cloud formation
in the Galactic halo,  usually in the context of 
high-velocity clouds, not gas in normal Galactic rotation 
\citep{shapiro, bregman80, ferrara94, wolfire95}.  
An exception is Houck and Bregman (1990), who 
explicitly studied the formation of corotating neutral clouds cooling from 
a low-temperature Galactic fountain. Some of their models 
produce clouds with a density and distance from the 
plane in the range of those found here, but with much smaller mass.  
If the clouds have formed in a fountain and  are falling to the plane, some
deviation from circular rotation might be observed, along with 
a projected component of the vertical velocity \citep{benjamin97}.  There is 
no evidence of a decrease in the average LSR velocity of the clouds 
with $|z|$ to at least 1 kpc, but any clouds which lag behind 
Galactic rotation, and whose line-of-sight 
velocity is thus below $V_t cos(b)$, would not have been examined in this
 study.  
Global models of the disk and halo (e.g., \citet{rosenbregman95,deAvillez00,
deAvillez01}) may soon be refined enough to allow a detailed 
comparison with the data. 
 Further observations, now in progress, should clarify 
the properties of the halo clouds, 
establish their distribution throughout the 
Galaxy and  relationship to other species, 
and reveal their internal structure.

\acknowledgments

I began this line of research some years ago in response to 
 a provocative question posed by Renzo Sancisi.  Thanks.  
  I also thank Jim Braatz and 
Ron Maddalena for help with observations and data reduction, and 
Bob Benjamin, John Dickey, and Harvey Liszt 
for discussion and comments on the manuscript.

\clearpage

% the tables

\input{tab1.tex}

% the figures

\clearpage
\begin{figure}
\epsscale{0.65}
\plotone{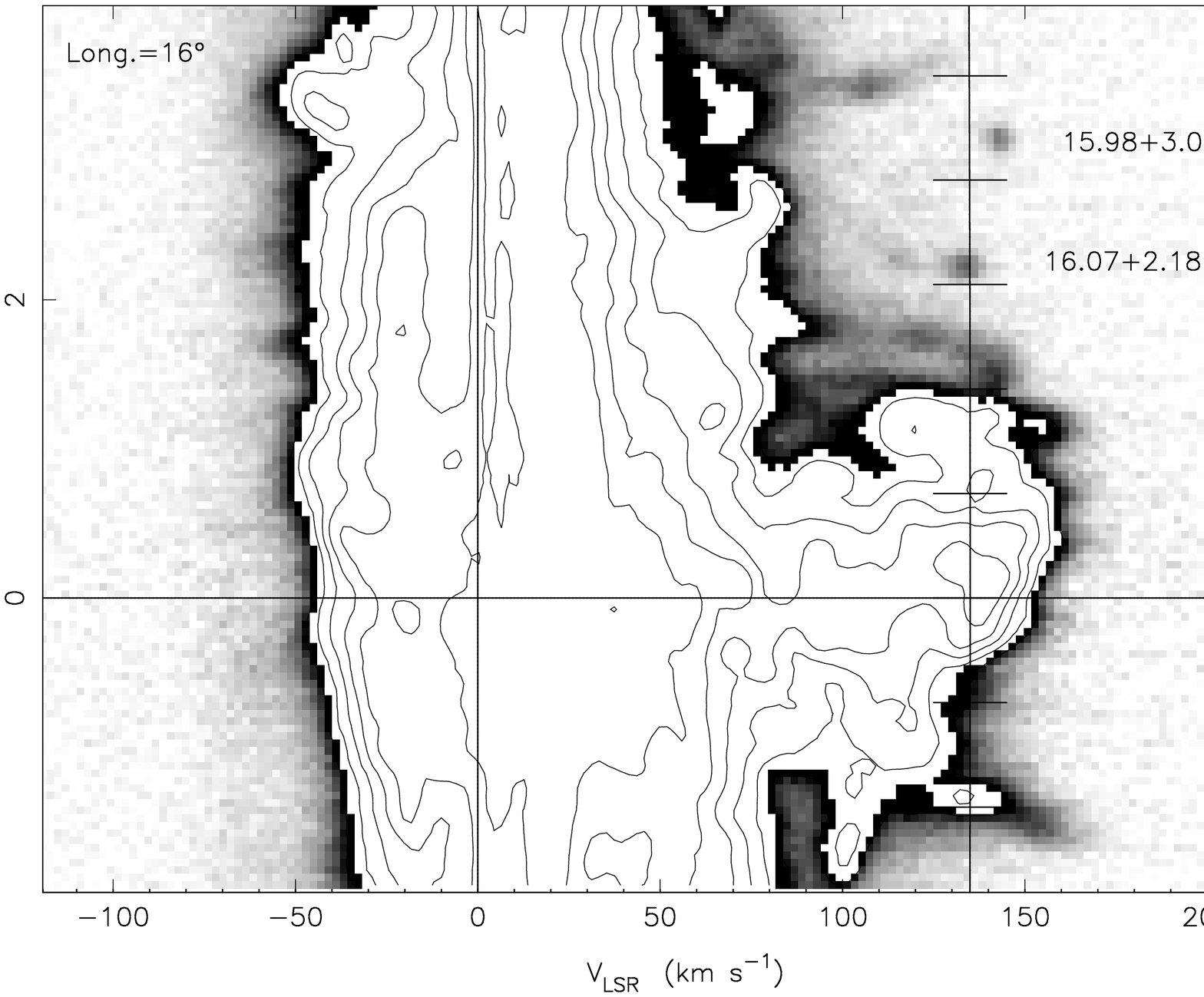}
\caption{The 21cm emission measured with the GBT every $3\arcmin$ 
in Galactic latitude  at $\ell = 16\arcdeg$. Contours are in factors of two
from the peak, and  a grey-scale is used for 
emission below 2 K.  The rms noise is typically 
$<0.1$ K.  The vertical line at +135 km s$^{-1}$ marks the 
terminal velocity from Galactic rotation as derived from 
$^{12}CO$ surveys.  Horizontal tics at the terminal velocity are located 
every 100 pc in distance from the Galactic plane.  Two small clouds
at the terminal velocity are identified by their coordinates as 
established from other observations.  From their velocity, they 
 must be quite near the tangent point 
8.2 kpc from the Sun, and thus lie 310 and 430 pc, 
respectively,  above the Galactic plane.
}
\end{figure}

\clearpage
\begin{figure}
\epsscale{0.65}
\plotone{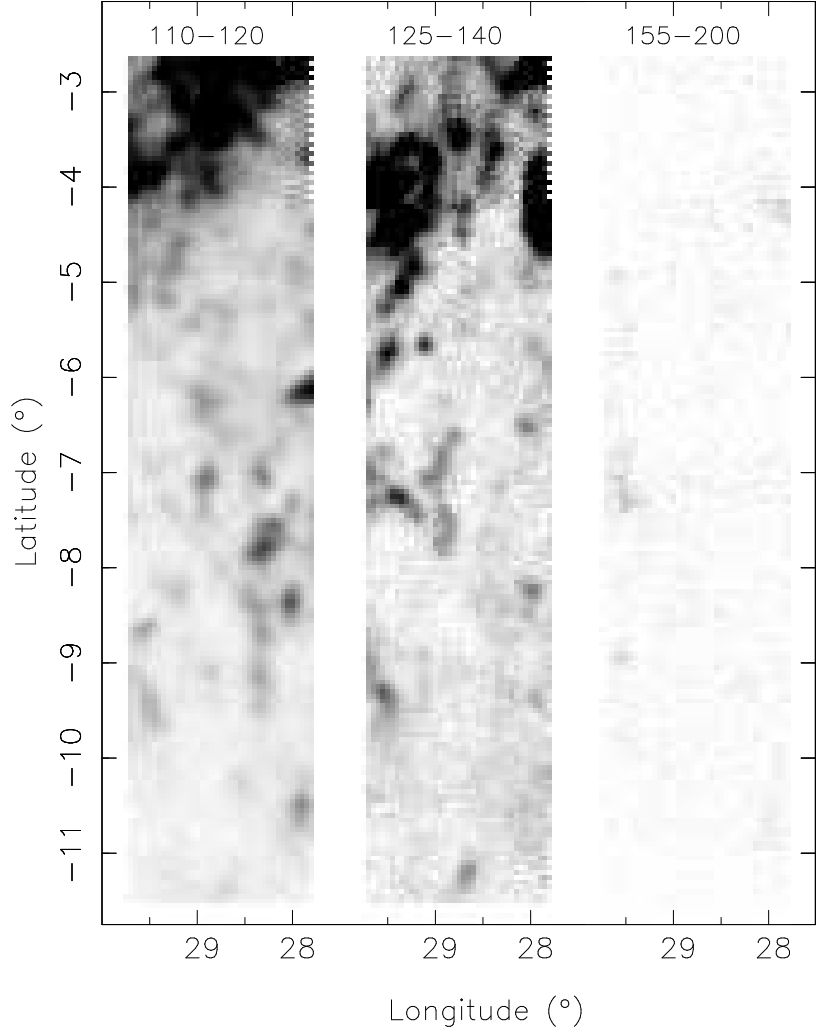}
\caption{GBT observations of the 21cm emission 
below the plane near $\ell = 29\arcdeg$ showing the 
cloudy structure of the \hi\ halo in this area.  
Spectra were taken at  $3\arcmin$ spacing in both coordinates; 
 each pixel was observed for a total of 10-20 seconds.  
At these longitudes the terminal velocity from Galactic rotation is 
105 km s$^{-1}$. Each panel shows \hi\ integrated over the 
velocity range given above it.  The left panel is clipped at 
$N_{HI} = 2 \times 10^{19}$ cm$^{-2}$ and the other two panels at 
half this value.  One degree in 
longitude or latitude in these maps  corresponds to a distance of 130 pc 
at the tangent point.  
Corotating halo clouds are found to $b = -11\fdg3$, or $z = -1500$ pc from 
the plane.  The virtual absence of clouds at the extreme velocities 
shows that the velocity of the clouds is derived mainly from Galactic 
rotation.  
 }
\end{figure}

\clearpage
\begin{figure}
\epsscale{0.65}
\plotone{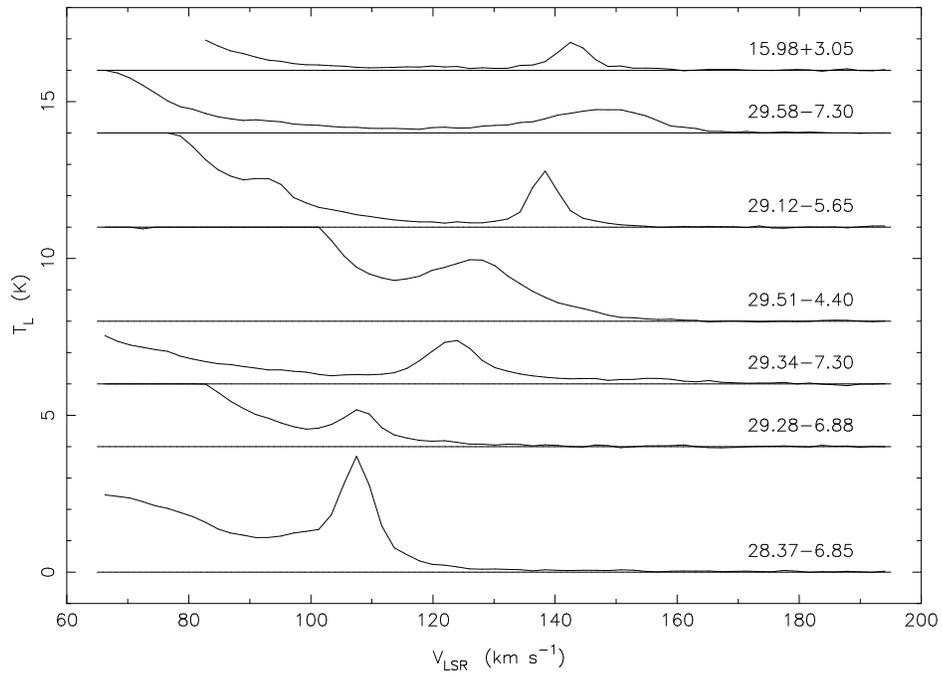}
\caption{\hi\  21cm spectra toward the center of representative 
bright halo clouds, identified by their Galactic coordinates. 
The uppermost spectrum is from a cloud shown in Figure 1, 
while the others are from  clouds in the field of Figure 2.  
The highest velocity peak in each spectrum  comes from the cloud. 
 }
\end{figure}

\end{document}

%% file: tab1.tex
%% Table 1 11 September 2002
\begin{deluxetable}{lcc}
\tablecolumns{3}
\tablenum{1}
\tablewidth{0pt}
\tablecaption{Halo Clouds: $\ell \approx 29\arcdeg, 
-12\arcdeg < b < -4\arcdeg$}
\tablehead{
\colhead{Property} & \colhead{Median} & \colhead{90\% Range} \\
\colhead{(1)} &\colhead{(2)} &\colhead{(3)} 
}  
\startdata 
 $V_{LSR}$ & $+118$ & $+105 \rightarrow +145 $\\
 $z$ (pc)   &   $-940$  & $-640 \rightarrow  -1210$ \\
 $T_L$ (K) & 1.0 & $0.4 \rightarrow 2.7$ \\
 $\Delta v$  (km s$^{-1}$)   &   12.2  & $  5.4 \rightarrow 26.3 $ \\
 $N_{HI}$  ($10^{19}$ cm$^{-2}$) &   $2   $ & 
 	$ 0.7 \rightarrow 6.3 $ \\
 Diameter (pc)    &  24  & $< 19  \rightarrow 35$ \\ 
 $\langle n \rangle $ (cm$^{-3})$ & 0.25 & $0.1 \rightarrow 0.9 $\\
 $M_{HI}$ ($M_{\sun}$) & 50 & $12 \rightarrow 290 $\\
\enddata
\end{deluxetable}

\vfill \eject